\documentclass{PoS}

\title{
\begin{picture}(0,0)(0,0)%
   \put(330,50){\makebox(0,0)[l]{\textnormal
{\normalsize LLNL-PROC-679808
}
}}%
\end{picture}%
SU(3) gauge theory with four degenerate fundamental fermions on the lattice}

\ShortTitle{SU(3) gauge theory with four degenerate fundamental fermions on the lattice}

\author{  Yasumichi Aoki$^a$,
  Tatsumi Aoyama$^a$,
  Ed Bennett$^{a b}$,
  \speaker{Masafumi Kurachi}$^c$\thanks{E-mail: kurachi@post.kek.jp},
  Toshihide Maskawa$^a$,
  Kohtaroh Miura$^{a d}$,
  Kei-ichi Nagai$^a$,
  Hiroshi Ohki$^e$, 
  Enrico Rinaldi$^f$,
  Akihiro Shibata$^g$,
  Koichi Yamawaki$^a$
  and 
  Takeshi Yamazaki$^h$ \\

  \hspace*{55mm} (LatKMI Collaboration) 
  \\
  
    $^a$
  Kobayashi-Maskawa Institute for the Origin of Particles and the Universe (KMI), Nagoya University, Nagoya 464-8602, Japan \\
  $^b$
  Department of Physics, Swansea University, Singleton Park, Swansea SA2 8PP, UK \\
  $^c$
  Institute of Particle and Nuclear studies, High Energy Accelerator Research Organization
(KEK), Tsukuba 305-0801, Japan\\
  $^c$ 
  Nuclear and Chemical Sciences Division, Lawrence Livermore National Laboratory, Livermore, CA 94550, USA \\
  $^d$
  Centre de Physique Theorique(CPT), Aix-Marseille University, Campus de Luminy, Case
907, 163 Avenue de Luminy, 13288 Marseille cedex 9, France\\
   $^e$
   RIKEN/BNL Research center, Brookhaven National Laboratory, Upton, NY, 11973, USA\\
   $^f$
Lawrence Livermore National Laboratory, Livermore, California, 94550, USA\\
   $^g$
  Computing Research Center, High Energy Accelerator Research Organization (KEK), Tsukuba 305-0801, Japan \\
  $^h$
  Graduate School of Pure and Applied Sciences, University of Tsukuba, Tsukuba, Ibaraki 305-8571, Japan
}

\abstract{The LatKMI Collaboration has been studying SU(3) gauge theories with a large number of fermion flavors, $N_f$. Here, we report results from lattice simulations of SU(3) gauge theory with four fundamental fermions. We first show the fermion mass dependence of $F_\pi$, $\langle\bar{\psi}\psi\rangle$ and their chiral extrapolations, showing evidence of chiral symmetry breaking. Then we report the mass spectrum of a vector meson and nucleon, showing that their behavior is very close to that of real-world QCD. We also show preliminary results of the measurement of the mass of the flavor-singlet scalar bound state. 
}

\FullConference{The 33rd International Symposium on Lattice Field Theory\\
		14 -18 July 2015\\
		Kobe International Conference Center, Kobe, Japan*}

\begin{document}
\section{Introduction}
As a part of the project studying large $N_f$ QCD, the LatKMI Collaboration has been investigating the SU(3) gauge theory with four fundamental fermions (four-flavor QCD). The main  purpose of studying four-flavor QCD is to provide a qualitative comparison to $N_f= 8$, $12$, $16$ QCD; however, a quantitative comparison to real-world QCD is also interesting. To make such comparisons more meaningful, it is desirable to use the same kind of lattice action consistently, so that qualitative difference of different theories are less affected by artifacts of lattice discretization. Here, we adopt the highly-improved staggered quark action with the tree-level Symanzik gauge action (HISQ/tree), which is exactly the same as the setup for our simulations for $SU(3)$ gauge theories with $N_f=8$, $12$ and $16$ fundamental fermions~\cite{Aoki:2013xza, Aoki:2012eq, Aoki:2014oma}. In the next section, we show the fermion mass dependence of $F_\pi$, $\langle\bar{\psi}\psi\rangle$, $M_\pi$, $M_\rho$, $M_N$ and their chiral extrapolations. In section 3, preliminary results of the measurement of the mass of the flavor-singlet scalar bound state will be reported.

\section{Chiral properties and masses of the vector meson and the nucleon}
For our study of four-flavor QCD, we generated gauge configurations at $\beta=3.7$ with three volumes, $(L, T) = (12, 18)$, $(16, 24)$, $(20, 30)$, and four or five bare mass parameter, $m_f$, in the range $0.01 \le m_f \le 0.05$ depending on the volume. 

\begin{figure}[b]
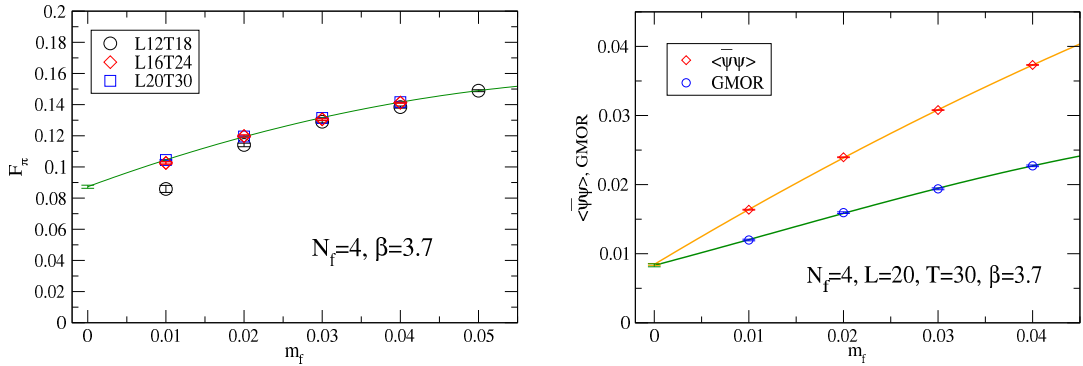
 
\begin{center} 
\includegraphics*[width=0.45\linewidth]{fpi_B3.7.eps} \ \ \ \ \ \ 
\includegraphics*[width=0.45\linewidth]{Nf4-PBP-GMOR.eps} 
\end{center}
\caption{ 
{\bf Left:} The decay constant of the NG boson, $F_\pi$, for various values 	of $m_f$ for $N_f=4$, $\beta=3.7$. The curve appearing in the figure is obtained by fitting the quadratic function to the largest volume data. 
{\bf Right:}  $\langle\bar{\psi}\psi\rangle$ for various values of $m_f$ for $N_f=4$, $\beta=3.7$. In the figure only the largest volume data is plotted. The chiral condensate estimated by using the GMOR relation is also plotted in the figure. The curves appearing in both figures are obtained by fitting the quadratic function to the largest volume data.}
\label{fig:Nf4-1}
\end{figure}

In the left panel of Fig.~\ref{fig:Nf4-1}, we show the decay constant of the Nambu--Goldstone (NG) boson, $F_\pi$, as a function of bare mass parameter $m_f$, and in the right panel, the chiral condensation, $\langle\bar{\psi}\psi\rangle$, is shown again as a functions of $m_f$.
Quadratic fits of $F_\pi$ and $\langle\bar{\psi}\psi\rangle$ show a non-zero value in the chiral limit, which can be considered as evidence for four-flavor QCD being in the chirally broken phase. As a consistency check, we also calculated the chiral condensate by using the GMOR relation, which is shown in the plot of $\langle\bar{\psi}\psi\rangle$. From the figure, we can see that the value of the chiral condensate obtained from the GMOR relation agrees with that of $\langle\bar{\psi}\psi\rangle$ in the chiral limit.

\begin{figure} 
\begin{center} 
\includegraphics*[width=0.5\linewidth]{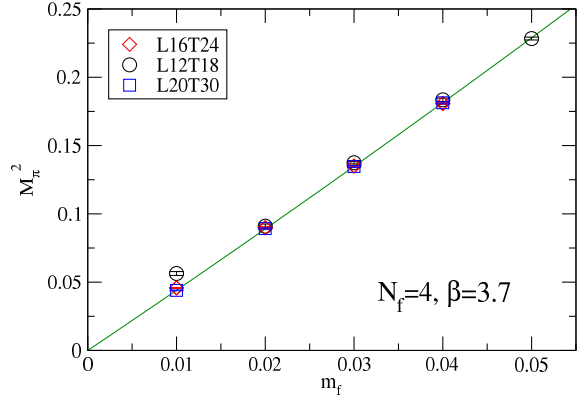} 
\end{center}
\caption{Mass-squared of the NG boson as a function of $m_f$. Quadratic fit to the largest volume data is also shown.}  
\label{fig:Nf4-2-1}
\end{figure}

\begin{figure} 
\begin{center} 
\includegraphics*[width=0.5\linewidth]{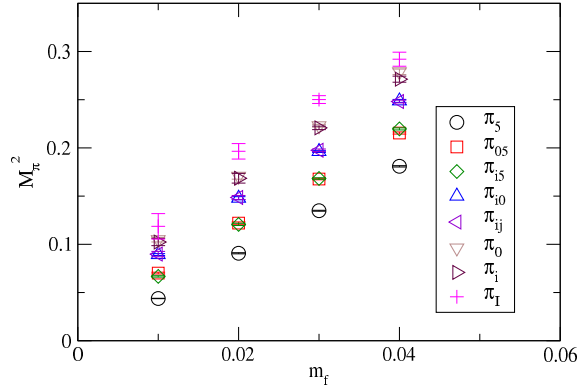}
\end{center}
\caption{Mass-squared of the NG boson for different tastes as functions of $m_f$.}  
\label{fig:Nf4-2-2}
\end{figure}

\begin{figure}[!h]
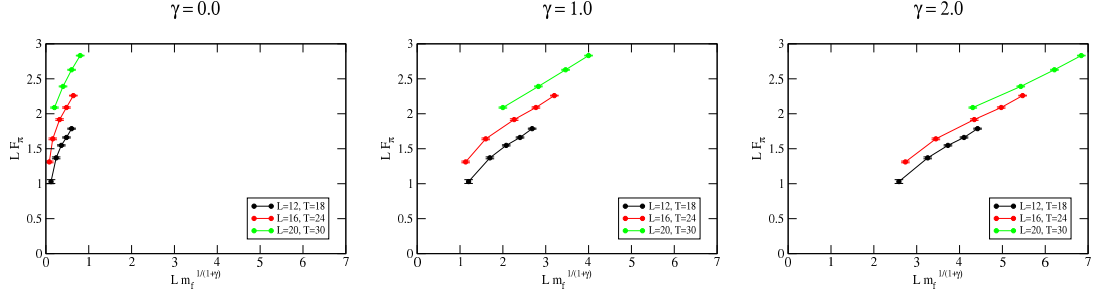
 
\makebox[.32\textwidth][r]{\includegraphics[width=.30\textwidth]{fpi_B3.7_g0.0.eps}}
\makebox[.32\textwidth][r]{\includegraphics[width=.30\textwidth]{fpi_B3.7_g1.0.eps}}
\makebox[.32\textwidth][r]{\includegraphics[width=.30\textwidth]{fpi_B3.7_g2.0.eps}}
\caption{
 Finite size hyperscaling test of $F_\pi$ in $N_f=4$ QCD. 
 Input values of $\gamma$ are, from left to right panels, $\gamma = 0.0, 1.0$ and $2.0$, respectively.
}
\label{fig:nf4fpihs}
\end{figure}

In Fig.~\ref{fig:Nf4-2-1}, we show the plot of the mass-squared ($M_\pi^2$) of the NG boson obtained from our simulations as a function of $m_f$. The largest volume data of $M_\pi^2$ can be fitted well by a liner ansatz; if the quadratic term is allowed, its coefficient is small. This is consistent with leading order predictions from chiral perturbation theory. In Fig.~\ref{fig:Nf4-2-2}, the mass-squared of the pseudoscalar mesons for different tastes are plotted as functions of $m_f$. One can see the almost constant shift in mass-squared for different tastes, which is consistent with the picture of the staggered $\chi$PT. (See also Ref.~\cite{Fodor:2009wk}) From these results, we conclude that the SU(3) gauge theory with four fundamental fermions is in the chiral symmetry breaking phase.
We also show the finite-size hyperscaling test \cite{DelDebbio:2010ze} for $N_f=4$ QCD 
by using the data of $F_\pi$ obtained here. In Fig.~\ref{fig:nf4fpihs}, 
we show the finite-size hyperscaling plot for input values of $\gamma= 0.0, 1.0$ and $2.0$. 
%
As we expect, the data show no alignment in the range of $0 \leq \gamma \leq 2$. 
This should be regarded as a typical property of QCD-like theory, and contrasted to the case of $N_f=8, 12$.

\begin{figure}[h] 
\begin{center} 
\includegraphics*[width=0.5\linewidth]{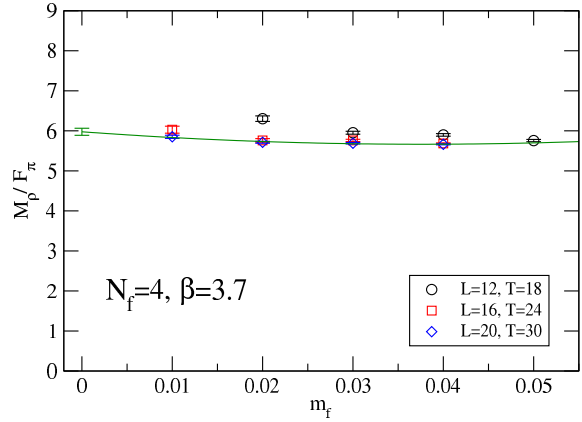}
\end{center}
\caption{The ratio of the mass of the vector meson, $M_\rho$, to $F_\pi$ for various values of $m_f$. Quadratic fit to the largest volume data is also shown.}  
\label{fig:Nf4-3-1}
\end{figure}
\begin{figure}[h] 
\begin{center} 
\includegraphics*[width=0.5\linewidth]{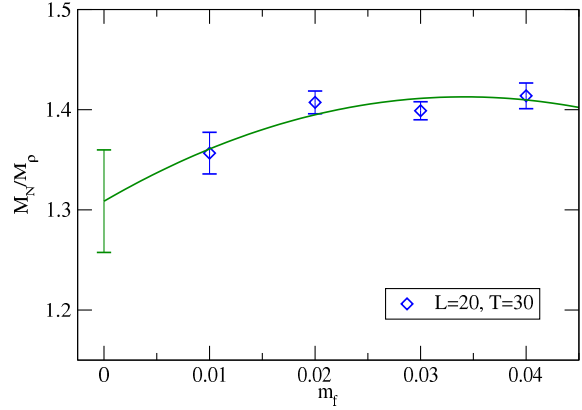}
\end{center}
\caption{The ratio of the mass of the nucleon, $M_N$, to $M_\rho$ for various values of $m_f$. Quadratic fit to the largest volume data is also shown.}  
\label{fig:Nf4-3-2}
\end{figure}
Fig.~\ref{fig:Nf4-3-1} shows the ratio of the mass of the vector meson, $M_\rho$, to $F_\pi$ for various values of $m_f$. The curve in the figure is the quadratic fit to the data of the largest volume, from which we can see the value of the ratio in the chiral limit being $M_\rho/F_\pi \simeq 6$. This value is very close to the ratio in the real-world QCD, $775\, {\rm MeV}/130\, {\rm MeV}\simeq 6$. In Fig.~\ref{fig:Nf4-3-2}, we show the ratio of the mass of the nucleon, $M_N$, to $M_\rho$ for various values of $m_f$ which are obtained from the largest volume data. The curve in the figure is quadratic fit to the data, from which we can see that the value of the ratio is, again, very close to that of the real-world QCD.

\begin{figure}[h] 
\begin{center} 
\includegraphics*[width=0.5\linewidth]{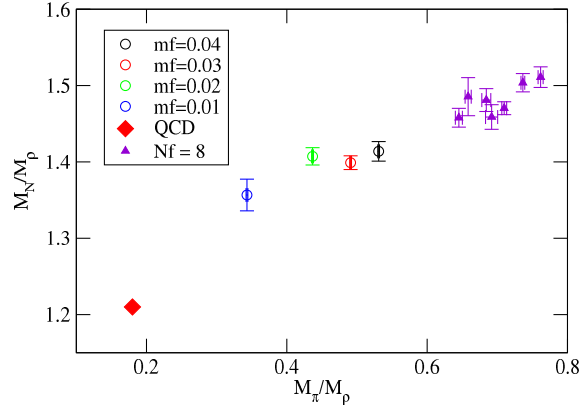}
\end{center}
\caption{Plot of $M_N/M_\rho$ versus $M_\pi/M_\rho$. The corresponding value of real-life QCD is indicated as diamond. For comparison, data of $N_f=8$ is plotted in the figure as filled triangle symbols. From top right to bottom left, the input bare mass is $m_f=$ 0.08, 0.06, 0.04, 0.03, 0.02, 0.015, and 0.012. }
\label{fig:Nf4-4-1}
\end{figure}
\begin{figure}[h] 
\begin{center} 
\includegraphics*[width=0.5\linewidth]{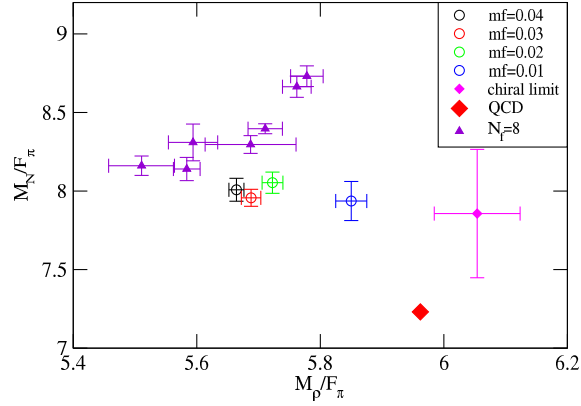}
\end{center}
\caption{Plot of $M_N/F_\pi$ versus $M_\rho/F_\pi$. The corresponding value in the chiral limit is indicated as magenta diamond and that of real-life QCD is indicated as red diamond. For comparison, the corresponding data of $N_f=8$ is plotted in the figure as triangle symbols. From top right to bottom left, the input bare mass is $m_f=$ 0.08, 0.06, 0.04, 0.03, 0.015, 0.012, 0.02. }
\label{fig:Nf4-4-2}
\end{figure}
In Fig.~\ref{fig:Nf4-4-1}, $M_N/M_\rho$ for various values of $M_\pi/M_\rho$ are plotted.  The corresponding value of real-life QCD is indicated by a diamond. For comparison, data of $N_f=8$ is plotted in the figure as filled circle symbols. From top right to bottom left of the $N_f=8$ data, the input bare mass is $m_f=$ 0.08, 0.06, 0.04, 0.03, 0.02, 0.015, and 0.012. In Fig.~\ref{fig:Nf4-4-2}, plot of $M_N/F_\pi$ for various values of $M_\rho/F_\pi$ are plotted. The corresponding value in the chiral limit is indicated as magenta diamond and that of real-life QCD is indicated as red diamond. For comparison, corresponding data of $N_f=8$ is plotted in the figure as triangle symbols. From top right to bottom left of the $N_f=8$ data, the input bare mass is $m_f=$ 0.08, 0.06, 0.04, 0.03, 0.015, 0.012, 0.02. From these figures, one can see that the data of $N_f =4$ is approaching to the real-world QCD in the chiral limit, while $N_f=8$ data stays in the area of the parameter space which is far from the real-world QCD value.

\section{Flavor-singlet scalar bound state}
\begin{figure}[t] 
\begin{center} 
\includegraphics[width=0.53\linewidth]{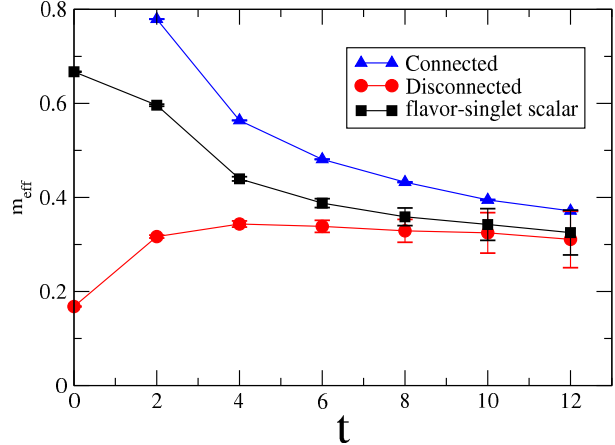} 
\end{center}
\caption{ 
The effective mass plot of the flavor-singlet scalar for $L=20$, $T=30$, $m_f=0.01$. The (black) square symbols represent effective mass calculated from the full scalar correlator, while (blue) triangle and (red) circle symbols represent those obtained from the connected and disconnected correlators, respectively.}
\label{fig:Nf4-5}
\end{figure}

The flavor-singlet scalar meson is a very interesting object both in large $N_f$ QCD and in real-world QCD. In the former case, it could be related to the 126 GeV Higgs boson discovered at the LHC, while for the latter case, it is related to the $\sigma$ meson. We calculate the flavor-singlet scalar correlator to extract the bound state mass by using the same technique we adopted for $N_f=8$ and $12$ QCD. (See Refs.~\cite{Aoki:2013zsa, Aoki:2014oha} for details. See also Ref.~\cite{Kunihiro:2003yj} for earlier lattice study on $\sigma$ meson.) The preliminary effective mass plot $L=20$, $T=30$, $m_f=0.01$ is shown in Fig.~\ref{fig:Nf4-5}.
The plateau is observed in the disconnected channel, and the mass extracted from it is about $0.3$, which is heavier than the pseudo NG boson ($M_\pi \sim 0.2$) in the same parameter. This is quite different from $N_f=8$ and $12$ QCD, in which it was shown that the mass of the flavor-singlet scalar bound state is comparable to or smaller than the pseudo NG boson~\cite{Aoki:2013zsa, Aoki:2014oha}.


\section{Summary}
In this proceedings, we showed our lattice simulation results for four-flavor QCD, which was performed by using the HISQ action with the tree-level Symanzik gauge action. It was shown that the theory is in the chiral symmetry breaking phase, and the ratio of $M_\rho$ to $F_\pi$, as well as  $M_N$ to $M_\rho$ in the chiral limit are quite close to that of real-world QCD.  Preliminary results for flavor-singlet bound state mass were also shown, and the qualitative difference between four-flavor QCD and (near-)conformal theories such as $N_f=8$ and $12$ QCD was pointed out. More detailed analysis with higher statistics and similar investigation in multiple mass parameter would help to understand the physical interpretation of the result, which will be pursued in the future.

\section*{Acknowledgments}
Numerical computations have been carried out on the high-performance computing systems at KMI ($\varphi$) and at the Research Institute for Information Technology in Kyushu University (CX400 and HA8000). This work is supported by the JSPS Grant-in-Aid for Scientific Research (S) No.22224003, (C) No.23540300 (K.Y.), for Young Scientists (B) No.25800139 (H.O.) and No.25800138 (T.Y.), and also by the MEXT Grants-in-Aid for Scientific Research on Innovative Areas No.23105708 (T.Y.), No. 25105011 (M.K.). E.R. acknowledges the support of the U.S. Department of Energy under Contract DE-AC52- 07NA27344 (LLNL). The work of H.O. is supported by the RIKEN Special Postdoctoral Researcher program.

\end{document}